\documentclass[12pt]{article}
\usepackage{graphicx}
\usepackage{xcolor}
\usepackage[hidelinks]{hyperref}
\hypersetup{
    colorlinks,
    linkcolor={red!50!black},
    citecolor={blue!50!black},
    urlcolor={blue!80!black}
}
\usepackage{amsmath}
\usepackage{cleveref}
\usepackage{wrapfig}

\newcommand\pubdate{\today}

\newcommand{\orcid}[1]{\,\href{https://orcid.org/#1}{\includegraphics[width=9pt]{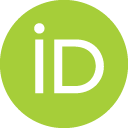}}}
\newcommand{\orcidTJ}{0000-0002-1334-7607} %
\newcommand{\orcidMK}{0000-0002-4665-3088} %
\newcommand{\orcidKK}{0000-0003-1412-447X} %
\newcommand{\orcidFO}{0000-0001-6799-2436} %
\newcommand{\orcidIS}{0000-0003-0373-474X} %
\newcommand{\orcidPR}{0000-0002-8570-5506} %
\newcommand{\orcidVB}{0000-0003-0148-0272}
\def\smu{{Department of Physics, Southern Methodist University,
    Dallas, TX 75275-0175, U.S.A.}}
\def\muenster{{Institut f{\"u}r Theoretische Physik, Westf{\"a}lische Wilhelms-Universit{\"a}t M{\"u}nster,
  \\Wilhelm-Klemm-Stra{\ss}e 9, D-48149 M{\"u}nster, Germany}}
\def\lpsc{{Laboratoire de Physique Subatomique et de Cosmologie, Université Grenoble-Alpes, 
    \\CNRS/IN2P3, 53 avenue des Martyrs, 38026 Grenoble, France}}
\def\IRFU{{IRFU, CEA, Université Paris-Saclay, 91191 Gif-sur-Yvette, France}}

\def\Title#1{\begin{center} {\Large #1 } \end{center}}
\def\Author#1{\begin{center}{ \sc #1} \end{center}}
\def\Address#1{\begin{center}{ \it #1} \end{center}}

\newcommand\pubblock{\rightline{\begin{tabular}{l}  \\ 
         \pubdate  \end{tabular}}}
\newenvironment{Abstract}{\begin{quotation}  }{\end{quotation}}
\newenvironment{Presented}{\begin{quotation} \begin{center} 
             PRESENTED AT\end{center}\bigskip 
      \begin{center}\begin{large}}{\end{large}\end{center} \end{quotation}}

\begin{document}
\begin{titlepage}
 \pubblock
\vfill
\Title{Fast evaluation of heavy-quark contributions to DIS in APFEL++}
\vfill

\Author{P. Risse$^{a,\dag}$\orcid{\orcidPR}, V. Bertone$^b$\orcid{\orcidVB}, T.~Je\v{z}o$^a$\orcid{\orcidTJ}, M.~Klasen$^a$\orcid{\orcidMK}, K.~Kova\v{r}\'{i}k$^a$\orcid{\orcidKK}, F.I.~Olness$^c$\orcid{\orcidFO}, I.~Schienbein$^d$\orcid{\orcidIS}}
\Address{\scriptsize $^a$\muenster \\ $^b$\IRFU \\ $^c$\smu \\$^d$\lpsc} 
\vfill
\begin{Abstract}
Mass-dependent quark contributions are of great importance to DIS processes. The simplified-ACOT-scheme includes these effects over a wide range of momentum transfers up to next-to-leading order in QCD. In recent years an improvement in the case of neutral current DIS has been achieved by using zero-mass contributions up to next-to-next-to-leading order (NNLO) with massive phase-space constraints. In this work, we extend this approach to the case of charged current DIS and provide an implementation in the open-source code \texttt{APFEL++}. The increased precision will be valuable for ongoing and future neutrino programs, the Electron-Ion-Collider and the studies of partonic substructure of hadrons and nuclei. A highly efficient implementation using gridding techniques extends the applicability of the code to the determination of parton distribution functions (PDFs).
\end{Abstract}
\vfill
\let\thefootnote\relax\footnotetext{$^\dag$\texttt{risse.p@uni-muenster.de}}
\begin{Presented}
DIS2023: XXX International Workshop on Deep-Inelastic Scattering and
Related Subjects, \\
Michigan State University, USA, 27-31 March 2023 \\
     \includegraphics[width=9cm]{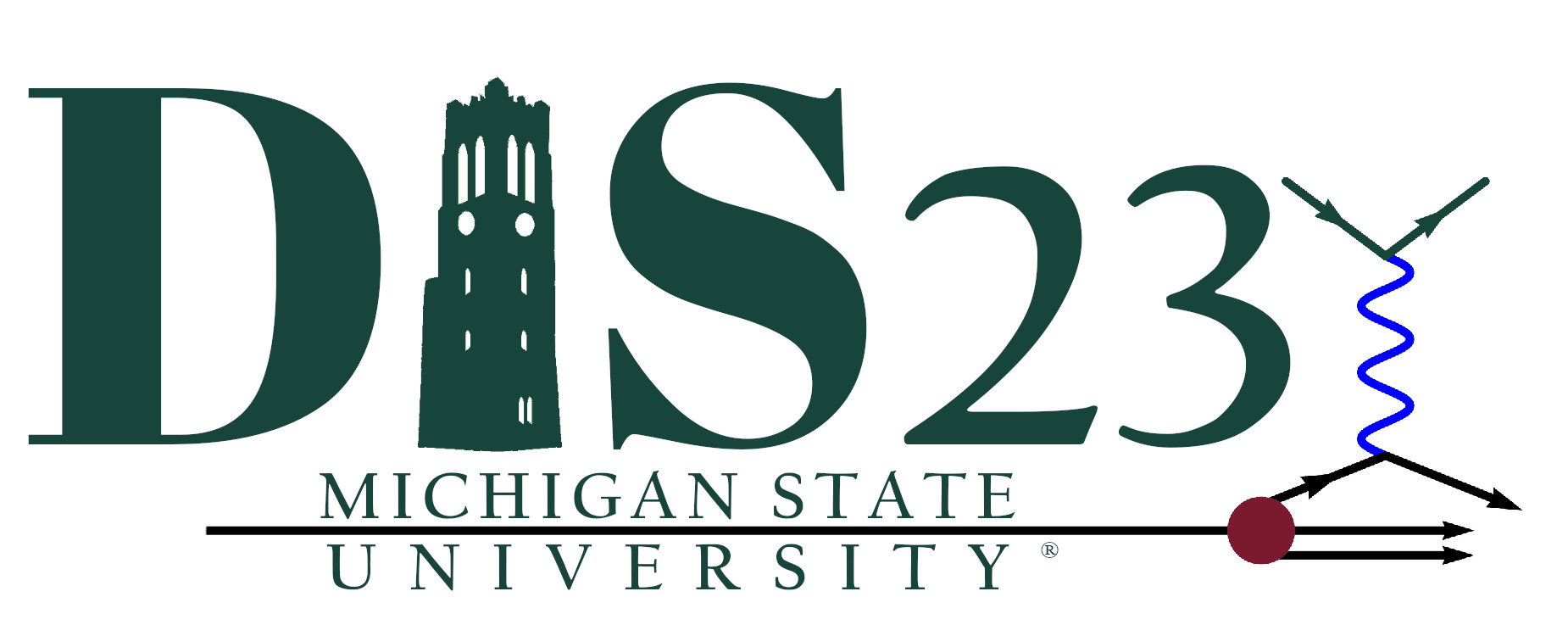}
\end{Presented}
\vfill
\end{titlepage}

\section{Motivation}
The production of heavy quarks in high energy processes is an important subject of study both theoretically and experimentally. The theory of heavy quark production in perturbative Quantum Chromodynamics (pQCD) is challenging due to the issues arising from an additional heavy quark mass scale. An accurate description must take the role of the heavy quark into account in the threshold region where the quark behaves like a typical heavy particle, and in the asymptotic region where the quark behaves like a parton similar to the light quarks $\{u,d,s\}$. The heavy quarks contribute up to 30\% to the structure functions in the kinematical region where the very precise DIS cross section data of HERA are located. These data form the backbone of modern global analyses of Parton Distribution Functions (PDFs) and it is therefore crucial to extend the heavy quark calculation schemes to higher orders in $\alpha_S$ for a precise extraction of the PDFs.

\section{The ACOT mass scheme}
One way of incorporating the heavy quarks is the zero-mass variable flavor number scheme (ZM-VFNS). It includes heavy quarks as a parton if the process energy $Q$ is far above the mass threshold, but calculates a structure function neglecting all mass terms in the Wilson coefficients. As $Q$ surpasses another threshold the next quark is added hence the name `variable flavor number scheme'. This is a good approximation at high values of $Q$ whenever the heavy quark acts like a parton, i.e. as a light quark. On the other hand the fixed flavor number scheme (FFNS) treats the heaviest quark as extrinsic to the hadron but the Wilson coefficient is fully massive. This returns accurate results close to the threshold but becomes increasingly inaccurate with rising process energy. 

Finally the ACOT scheme \cite{Aivazis:1993pi} provides a mechanism to incorporate the heavy quark into the calculation over a wide range of $Q$. At NLO its key ingredient is the so-called subtraction term which removes the double counting where the combination of the leading order part and the DGLAP evolution overlap with the NLO contribution. This allows for the ACOT scheme to reduce to the appropriate limit both as \mbox{$m/Q\rightarrow 0$} and \mbox{$m/Q \rightarrow \infty$}: In the high energy limit it reduces exactly to the \mbox{ZM-VFNS}. Whenever $m/Q\sim 1$ the heavy quark decouples from the PDFs and therefore the ACOT scheme reduces to the FFNS for $m/Q\gg 1$.
\begin{figure}
    \centering
    \includegraphics[width=0.7\textwidth]{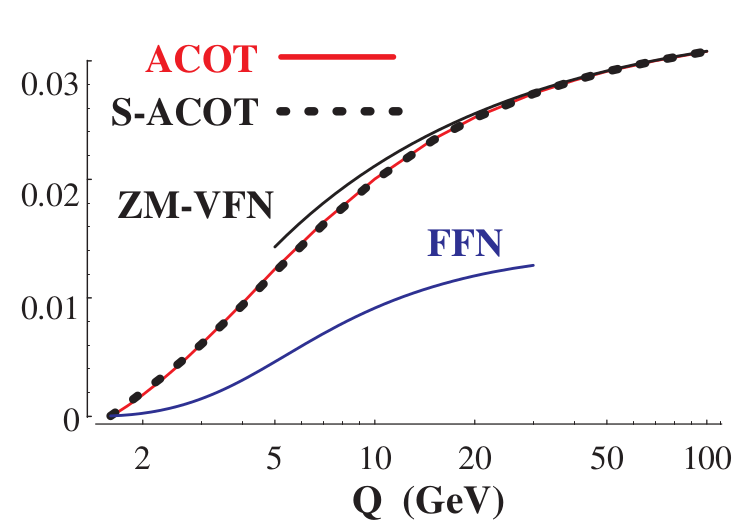}
    \caption{The charm contribution to the $F_2$ structure function at $x=0.1$ for NLO DIS as a function of $Q$. The ACOT (red) and S-ACOT (black, dashed) are virtually identical and interpolate between the ZM-VFNS (black) and the FFNS (blue). Figure taken from \cite{Stavreva:2012bs}.}
    \label{fig:ACOT_FFN_ZM-VFNS}
\end{figure}
Fig.\,\ref{fig:ACOT_FFN_ZM-VFNS} displays the charm contribution to the $F_2$ structure function for the ACOT scheme (red), the \mbox{ZM-VFNS} (black) and the FFNS (blue). The ACOT scheme becomes identical with the \mbox{ZM-VFNS} at high energies and reduces to the FFNS at the charm threshold. The dashed black line is the Simplified-ACOT (S-ACOT), a variant (not an approximation) of the ACOT scheme, where in certain parts of the calculation the mass of the quark can be set to zero. Both schemes are virtually identical.

The study in Ref.\,\cite{Stavreva:2012bs} (and the references therein for the detailed calculations) extended the original formulation of the ACOT scheme to next-to-next-to leading order (NNLO) in QCD. They concentrated on the $F_2$ and $F_L$ structure function in the neural current case. The key idea was the observation that the dominant mass contribution to the structure function was in fact a result of the phase space constraint. They defined a general scaling variable $\chi(n)$ as 
\begin{equation}
    \chi(n) = x\left[1+\left(\frac{n\,m_H}{Q}\right)^2\right]
\end{equation}
where $m_H$ is the heavy quark's mass and $n=\{0,1,2\}$. Here $n=0$ corresponds to the massless result and $n=1,2$ to having one or two massive quarks in the Feynman graphs. In the proposal, they use the full ACOT scheme for the LO and NLO contribution and combine them with the zero mass coefficients at NNLO, i.e. remove the dynamic mass but apply the massive $\chi(n)$-rescaling. For that, the computation at NNLO needs to be dissected into the individual flavor contributions to apply the $\chi(n)$ with the proper mass. The final result shows only a small dependence on $n$, suggesting only a small error is made in the approximation, as the dominant contribution from the mass is in the phase space constraint.

\section{Extension to the charged current case}
The ansatz from the neutral current case can be transferred to the charged current; only a few differences demand consideration. The individual flavor contributions are to be treated carefully as there is a flavor change at the electroweak vertex. Since all possible flavor changes from the CKM-matrix are taken into account, terms with two different masses occur. In fact, similar considerations were needed in the neutral current for certain diagrams as a radiated gluon can produce a quark anti-quark pair of any flavor. The choice was to attribute these contributions to the heavier of the two quarks and apply the $\chi(n)$-rescaling correspondingly. We keep this choice in the charged current implementation.
\begin{figure}
  \begin{center}
    \includegraphics[width=0.58\textwidth]{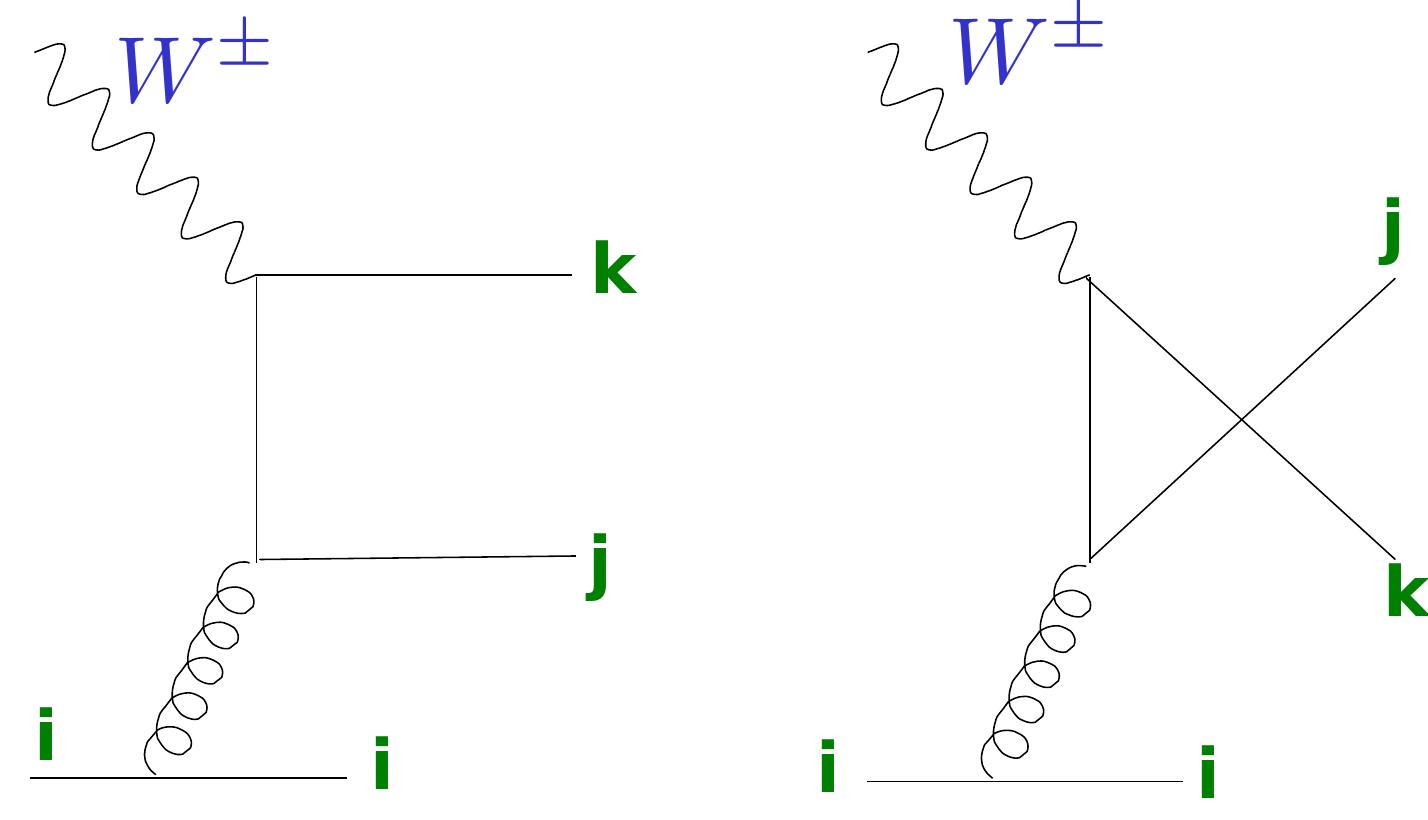}
  \end{center}
  \caption{Sample diagrams contributing with up to three different heavy quark masses to the charged current structure function.}
  \label{fig:diagram_three_masses}
\end{figure}
However, with the flavor change at the vertex three heavy quark masses can be present at the same time. Fig.\,\ref{fig:diagram_three_masses} displays two sample diagrams where the flavors $i,j$ and $k$ can be the charm, bottom and top quark. Again, we apply the $\chi(n)$-rescaling using the heaviest quark mass. Although this is certainly the edge case, we include $n=3$ in the spread of scalings. 

\begin{figure}
    \centering
    \includegraphics[width=0.5\textwidth]{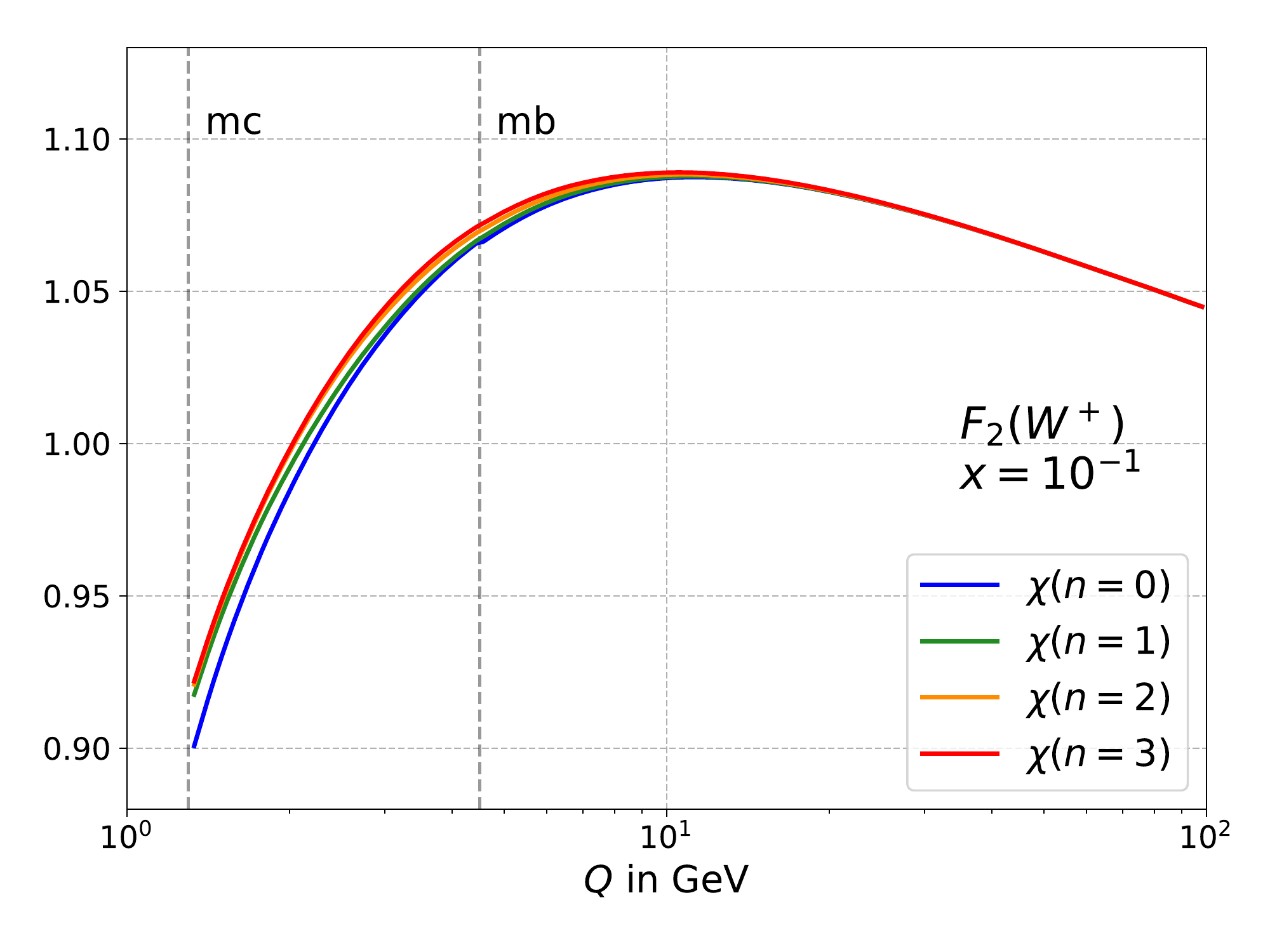}%
    \includegraphics[width=0.5\textwidth]{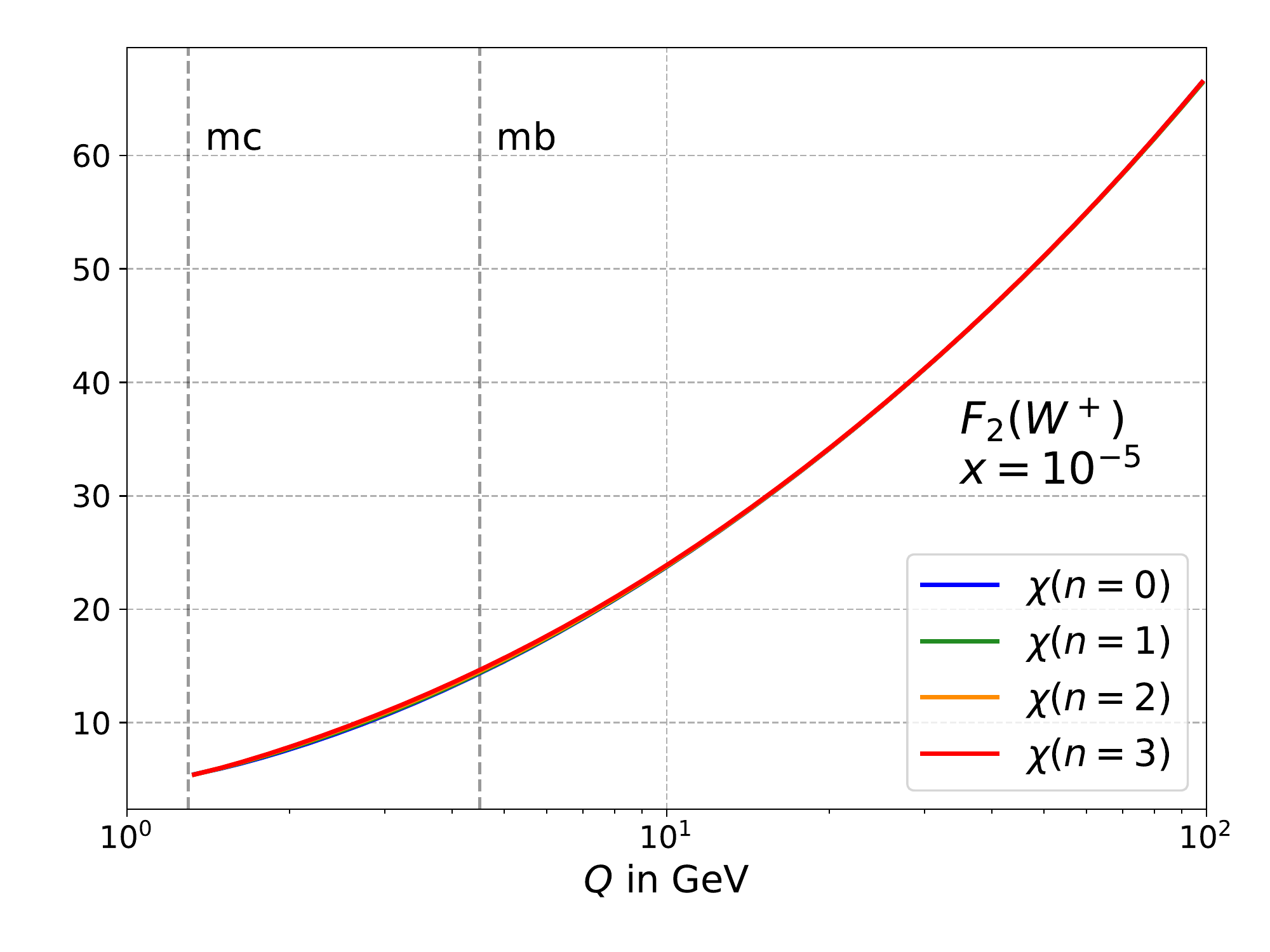}
    \caption{The $F_2$ structure function for a $W^+$ exchange as a function of $Q$. The left plot shows $F_2$ at $x=0.1$ and the right plot at $x=10^{-5}$ with color-coded $n$ values for the $\chi(n)$-rescaling. The structure function shows only a mild dependence on $n$. This result was obtained with the \texttt{APFEL++} code.}
    \label{fig:Wp_F2}
\end{figure}

The result for a $W^+$ exchange is shown in Fig\,\ref{fig:Wp_F2}. Here the $F_2$ structure function is displayed as a function of $Q$ for two $x$ values: $x=0.1$ (left) and $x=10^{-5}$ (right). With a color-coding of blue ($n=0$), green ($n=1$), yellow ($n=2$) and red ($n=3$) different values for $n$ in the $\chi(n)$-variable are compared. The charm and bottom thresholds are depicted as black dashed lines. In the left plot the structure function shows only a mild dependence on $n$ even for the edge case of $n=3$. The only outlier is the zero-mass result with $n=0$. The agreement improves towards lower $x$ values as to be seen in the right figure. 

\section{Numerical implementation in APFEL++}
The numerical implementation was incorporated in the open source code \texttt{APFEL++} \cite{Bertone:2017gds,Bertone:2013vaa}. The code is focused on a fast and memory efficient implementation of theory calculations in pQCD like solving the DGLAP evolution equations or calculating structure functions. The code is able to compute structure functions in the \mbox{ZM-VFNS} or in the FONLL-scheme \cite{Forte:2010ta} up to NNLO. 

A fast evaluation is achieved by precomputing the expensive part of the calculation: a structure function $F_\lambda$ is constructed by performing a Mellin convolution over a Wilson coefficient $C_\lambda(\xi)$ and a PDF $f(\xi)$ (we drop the summation over the PDF flavors here). However, the evaluation of a Wilson coefficient is computationally much more expensive than a PDF. The strategy in the code is to discretise the integration variable $\xi$ by means of a grid $g=\{\xi_0,\dots,\xi_N\}$ and interpolate the parton distribution using a moving set of interpolating functions:
\begin{equation}
    f(\xi) = \sum_\beta w_\beta(\xi)f_\beta,
    \label{eq:PDF_interpolation}
\end{equation}
where $f_\beta$ is the PDF evaluated at a grid node $\xi_\beta$. Plugging Eq.\,\ref{eq:PDF_interpolation} into the calculation of a structure function $F_\lambda(x)$ and assuming the value of $x$ to be equal the to $\alpha$-th grid node we find that the calculation is reduced into a simple matrix vector multiplication:
\begin{align}
    F_\lambda(x_\alpha) &= \int^1_{x_\alpha}\frac{\mathrm{d}\xi}{\xi}C_\lambda(\xi)f\left(\frac{x_\alpha}{\xi}\right) = \sum_\beta O_{\alpha\beta}f_\beta \\
    \text{with}\qquad O_{\alpha\beta} &= \int^1_{x_\alpha}\frac{\mathrm{d}\xi}{\xi}C_\lambda(\xi)w_\beta\left(\frac{x_\alpha}{\xi}\right).
\end{align}
Once all entries of $O_{\alpha\beta}$ have been precomputed the value of $F_\lambda$ for an arbitrary value of $x$ can be obtained through interpolation. A multiplication of a matrix and a vector is very fast and allows to recompute the structure function with different PDFs very efficiently, making the implementation especially well-suited for PDF fitting.

We implemented the S-ACOT mass scheme up to NLO (neutral current and charged current) in \texttt{APFEL++}, along with an implementation of approximative \mbox{S-ACOT} at NNLO for neutral current and charged current. The calculation of the structure function in Fig\,\ref{fig:Wp_F2} was obtained using \texttt{APFEL++}. The computational speedup in the neutral current compared to the original implementation is of an order \mbox{of $\mathcal{O}(100)$}.

\section{Conclusion}
In recent years, the ACOT mass scheme was brought up to NNLO by approximating the phase space constraints from a massive quark by the $\chi(n)$-rescaling and thereby capturing the dominant contribution from heavy quarks to DIS structure functions. We extended the approach to the charged current. The main task was to dissect the calculation of a structure function w.r.t. its individual quark contributions and applying the $\chi(n)$-rescaling appropriately. As in the neutral current case we find only a mild dependence on the scaling factor $n$. 

A numerical implementation was incorporated in the framework of the open source code \texttt{APFEL++}. The code provides a very fast implementation of the calculation of structure functions by the means of interpolation and is especially well-suited for PDF fitting. We implemented the S-ACOT scheme up to NLO, and the approximate S-ACOT at NNLO for neutral and charged currents. The neutral current implementation was checked against the original implementation and yields a computational speedup of up to $\mathcal{O}(100)$.

\footnotesize
\bibliographystyle{utphys}
\bibliography{references}
\typeout{get arXiv to do 4 passes: Label(s) may have changed. Rerun}
\end{document}